\documentclass[conference]{IEEEtran}
\IEEEoverridecommandlockouts
\usepackage{cite}
\usepackage{amsmath,amssymb,amsfonts}
\usepackage{algorithmic}
\usepackage{graphicx}
\usepackage{textcomp}
\usepackage{xcolor}
\usepackage[flushleft]{threeparttable}

\usepackage{amsmath,amssymb,amsfonts}
\usepackage{algorithmic}
\usepackage{graphicx}
\usepackage{textcomp}
\usepackage{xcolor}
\usepackage{multirow}

\usepackage{threeparttable}
\usepackage{tabularx}
\usepackage{siunitx}
\usepackage{amssymb}
\usepackage{pifont}
\newcommand{\cmark}{\ding{51}}%
\newcommand{\xmark}{\ding{55}}%
\usepackage{bm}

\def\BibTeX{{\rm B\kern-.05em{\sc i\kern-.025em b}\kern-.08em
    T\kern-.1667em\lower.7ex\hbox{E}\kern-.125emX}}

\makeatletter

\def\ps@IEEEtitlepagestyle{
  \def\@oddfoot{\mycopyrightnotice}
  \def\@evenfoot{}
}
\def\mycopyrightnotice{
  {\footnotesize
  \begin{minipage}{\textwidth}
  \centering
  ~\copyright~ 2023 IEEE.  Personal use of this material is permitted.  Permission from IEEE must be obtained for all other uses, in any current or future media, including reprinting/republishing this material for advertising or promotional purposes, creating new collective works, for resale or redistribution to servers or lists, or reuse of any copyrighted component of this work in other works.
  \end{minipage}
  }
}

\usepackage{fancyhdr}   

\fancypagestyle{firstpage}{     
  \fancyhf{}
  \chead{\textcolor{gray}{This article has been accepted for publication in the proceedings of the \\ 2023 IEEE international workshop on Metrology for Industry 4.0 \& IoT.}}
  \fancyfoot[C]{\small{\textcolor{gray}{~\copyright~ 2023 IEEE.  Personal use of this material is permitted.  Permission from IEEE must be obtained for all other uses, in any current or future media, including reprinting/republishing this material for advertising or promotional purposes, creating new collective works, for resale or redistribution to servers or lists, or reuse of any copyrighted component of this work in other works.}}}

}

\begin{document}

\title{Fully Automatic Gym Exercises Recording: An IoT Solution\\
}

\author{\IEEEauthorblockN{Sizhen Bian}
\IEEEauthorblockA{\textit{ETH Zürich} \\
\textit{PBL-DITET}\\
Zürich, Switzerland \\
sizhen.bian@pbl.ee.ethz.de}
\and
\IEEEauthorblockN{Alexander Rupp}
\IEEEauthorblockA{\textit{ETH Zürich} \\
\textit{PBL-DITET}\\
Zürich, Switzerland\\
ruppa@student.ethz.ch}
\and
\IEEEauthorblockN{Michele Magno}
\IEEEauthorblockA{\textit{ETH Zürich} \\
\textit{PBL-DITET}\\
Zürich, Switzerland\\
michele.magno@pbl.ee.ethz.de}
}

\maketitle

\begin{abstract}
In recent years, working out in the gym has gotten increasingly more data-focused and many gym enthusiasts are recording their exercises to have a better overview of their historical gym activities and to make a better exercise plan for the future.
As a side effect, this recording process has led to a lot of time spent painstakingly operating these apps by plugging in used types of equipment and repetitions. This project aims to automate this process using an Internet of Things (IoT) approach. Specifically, beacons with embedded ultra-low-power inertial measurement units (IMUs) are attached to the types of equipment to recognize the usage and transmit the information to gym-goers and managers. We have created a small ecosystem composed of beacons, a gateway, smartwatches, android/iPhone applications, a firebase cloud server, and a dashboard, all communicating over a mixture of Bluetooth and Wifi to distribute collected data from machines to users and gym managers in a compact and meaningful way. The system we have implemented is a working prototype of a bigger end goal and is supposed to initialize progress toward a smarter, more efficient, and still privacy-respect gym environment in the future. A small-scale real-life test shows 94.6\% accuracy in user gym session recording, which can reach up to 100\% easily with a more suitable assembling of the beacons. This promising result shows the potential of a fully automatic exercise recording system, which enables comprehensive monitoring and analysis of the exercise sessions and frees the user from manual recording.  The estimated battery life of the beacon is 400 days with a 210 mAh coin battery. We also discussed the shortcoming of the current demonstration system and the future work for a reliable and ready-to-deploy automatic gym workout recording system.

\end{abstract}


\section{Introduction}

\thispagestyle{firstpage}

Regularly visiting the gym has been an important part of a healthy lifestyle for many people worldwide \cite{kercher20222022}. Thus, gym activity digitalization has become a popular topic in industry and academia. Progress in the "smartification" of gym exercising is currently moving on many different paths, and no solution has been widely and properly adopted. One can find many pieces of workout equipment, but those are usually still seen as gadgets used for more casual workout programs rather than for a fully functioning gym ecosystem. A fully automatic gym exercise tracking enables gym enthusiasts thoroughly focus on their workouts and frees them from swiping and tapping a commercial App like Bodyspace to record their workout history \cite{padmasekara2014fitness}.

\begin{table*}[!t]
\centering
\caption{Gym exercises recognition solutions}
\label{relatedwork}
\begin{threeparttable}
\begin{tabular}{ p{1.2cm} p{1.6cm}  p{1.6cm} p{1.9cm} p{1.0cm} p{2.0cm} p{2.5cm} p{2.2cm} }
\hline
authors/ year & Sensor & Cost(\$) & Power Consumption($\approx$)  & Privacy-respect &  Targeted Exercises & Accuracy & Computing model \\
\hline
\cite{hussain2022sensor}-2022 & Single accelerometer &  a few & mW & \cmark & free
weight exercises & 82\% for 42 exercises & LSTM\\
\hline
\cite{tian2021wearable}-2021 & Two IMUs & tens & mW & \cmark & full-body exercise  & 91.26\% for 8 exercises & fused ML models\\
\hline
\cite{koskimaki2017myogym}-2017 & IMU+ EMG &  Tens & mW & \cmark  &free-weight, arm-involved  & 71.6\% for 30 exercises & linear discriminant analysis \\
\hline
\cite{bian2022using}-2022  & Capacitive & a few & mW  & \cmark & leg-dominated machine-free & 89\% on 7 exercises & random forest  \\
\hline

\cite{bian2019passive}-2019 & Capacitive & a few & mW & \cmark & full-body exercise  & 63\% for 8 exercises  & TCN \\
\hline
\cite{bian2022exploring}-2022 & IMU+ Capacitive & a few & mW & \cmark & full-body exercise   & 90.4\% for 11 exercises  & Resnet \\
\hline
\cite{fu2018fitness}-2018 & Ultrasound  & no extra cost (smartphone) & unknow &\cmark  & bicycles, toe touches, squats & averaged 92\% for 3 exercises  & CNN \\
\hline
\cite{tiwari2021mmwave}-2021 & mmWave Radar  & tens & W &\cmark  & no limitation & averaged 93\% for 7 exercises  & CNN \\
\hline

\cite{radhakrishnan2021w8}-2021 & Accelero- \& Magnetometer  & a few & mW &\cmark  & weight stack-based exercises & 98.74\% for 14 exercises  & random forest \\
\hline

\cite{sundholm2014smart}-2014 & Textile Pressure& hundreds & mW  & \cmark & On-ground exercises & 82.5\% for 10 exercises & Dynamic Time Warping \\
\hline

\cite{ganesh2020personalized}-2020 & RGB Camera & hundreds & W & \xmark & no limitation & 98.5\% for 5 exercises & transfer learning \\
\hline
\cite{chen2022fitness}-2022 & RGB Camera & hundreds & W & \xmark & no limitation & 98.56\% for 12 exercises & MLP \\
\hline
\textbf{Our solution} & \textbf{Accelero- meter}   & \textbf{a few} & \textbf{uW} &  \cmark &  \textbf{equipment-involved}  & \textbf{100\% for all equipment exercises } & \textbf{threshold-based interrupt} \\
\hline

\hline
\end{tabular}
\end{threeparttable}
\end{table*}

Current solutions for automatic gym activity recording rely either on the applications installed on smart wearable devices, which require a lot of interaction from the user, or the embedded algorithms for automatic gym recognition based on the data collected from different sensors. Table \ref{relatedwork} listed the recent research work on gym exercise recognition sorted by the applied sensing modalities. Wearable solutions are inherently popular as no extra off-body devices are needed. Thus, the inertial measurement unit was widely explored as it is embedded in every portable smart device and enjoys the advantages of cost, power consumption, and privacy. In \cite{hussain2022sensor}, the authors used a single chest-mounted tri-axial accelerometer, followed by an LSTM neural network, to recognize a wide range of gym-based free-weight exercises. To be noticed, although the recognition accuracy is impressive (82\%) for 42 exercises, the data acquisition process involves only four athletes and the testing data set includes data from all four athletes. And the need for a rather large body harness for information gathering also limits the possibility of widespread use. Also, the chest-mounted sensor resulted in the misclassification of exercises that do not involve chest movement. Thus, to get better accuracy, extra inertial measurement units are needed to cover all the moving body parts during the exercises. Tian et al. \cite{tian2021wearable} analyzed the positions of the sensor on the body on the recognition performance, a stratification fusion method using two sensors was proved to efficiently identify eight kinds of gym exercises with an accuracy of 91.26\%. Besides the inertial sensors, the EMG \cite{koskimaki2017myogym} and passive capacitive sensors \cite{bian2019passive, bian2022using} were also explored in a wearable way for exercise recognition. Capacitive sensing was proved to be a competitive motion sensing approach for wearables enjoying the same advantage of low cost and low power consumption \cite{bian2022contribution}. With non-wearable sensors, including the image-based approaches, sensors are normally deployed near the user, either targeting the user for signal perceiving \cite{fu2018fitness, tiwari2021mmwave, sundholm2014smart} or the equipment \cite{radhakrishnan2021w8}. To be noticed, all the above-mentioned sensors for exercise recognition were explored on a very limited number of exercises. Large-scale exercise recognition is still challenging the efficiency of those sensing modalities. The image-based solution backed by the computer version technique could solve this problem as it perceives the full body movement through continuous frames. Thus, a near-perfect recognition accuracy \cite{ganesh2020personalized, chen2022fitness} could be achieved by using advanced machine learning skills. However, a few factors that the vision solution suffers from makes it hard for broad deployment, like privacy issues, as not only the data from user could be perceived and processed, and the high computing load caused by the large data volumes, etc.

This work addresses the above-described problem, aiming to provide gym enthusiasts with a fully automatic exercise recording system. More specifically, the paper proposes and presents an IoT system, including hardware, firmware, and cloud platform designed for equipment activity data sensing and distribution. On the hardware side, the inertial sensor-embedded, ultra-low-power beacons will be developed in advertising mode and deployed to the gym equipment, recognizing machine activity (the machine type and the repetition number) and transmitting the information via wireless packages. Nearby users could use their smartwatch or smartphone to scan the advertisements around and record the one with the strongest Bluetooth Received Signal Strength Indication (RSSI), which means this advertisement was generated by the moving equipment the user is working on. Meanwhile, a gateway will be deployed to listen to all the packages and send the processed data to a Firebase cloud server. Thus a dashboard-like interface can be presented to the gym manager for statistical data analysis, like everyday equipment usage situation, studio crowd analysis, etc. 

Compared with other solutions, this IoT approach enjoys the following advantages:
\begin{enumerate}
\item Low cost and low power consumption. Only the inertial sensor-embedded beacons are needed for equipment status data perceiving, and a very light computing load is needed for data processing. A coin battery could support over one year of life for the beacon.
\item Easy of deployment. Beacons are deployed onto the existing machines in a plug-and-play manner, making the IoT system economically deployable and easy to maintain. 
\item High accuracy. The broadcasting package includes both the equipment type (pre-configured) and the repetition number that can be reliably sensed by the inertial sensor in the beacon. 
\item Privacy-respect. Only the equipment status data is broadcasted. No personal user data is involved in the system. 
\item Large number of exercise types. Almost all types of exercises assisted by the equipment in a gym studio can be covered.
\item Extra benefits for gym managers. The equipment status advertisements broadcasted by the beacons are collected by the gateway and presented to the gym manager in the form of a cloud-based dashboard for potential services like crowd density monitoring, equipment maintenance, etc.
\end{enumerate}

The proposed IoT solution for fully automatic gym exercise recording aims to provide considerable improvements in the area as we're using considerably smaller and energy-efficient devices. Unlike current commercially available intelligent gym equipment with real-time data visualization and feedback, the proposed system is economical for integration into existing gym equipment. The digitalization of the gym studio with the IoT approach provides benefits for both users and managers, with which they could make more informed decisions.

In summary, we have the following two contributions:
\begin{enumerate}
\item the design and implementation of an IoT system for digitalizing the gym studio with minimum cost, providing fully automatic gym exercise recording for users and equipment usage status monitoring for managers.
\item With the real-life experiment, we showed the system's feasibility in digitalizing gym activities with near-perfect accuracy in exercise counting and information broadcasting. As a result, both the user and the manager get reliable real-time data of their interest.
\end{enumerate}

\section{System architecture}

\begin{figure}[!t]
\begin{minipage}[t]{1.0\linewidth}
\centering
\includegraphics[width=0.85\textwidth,height=4.0cm]{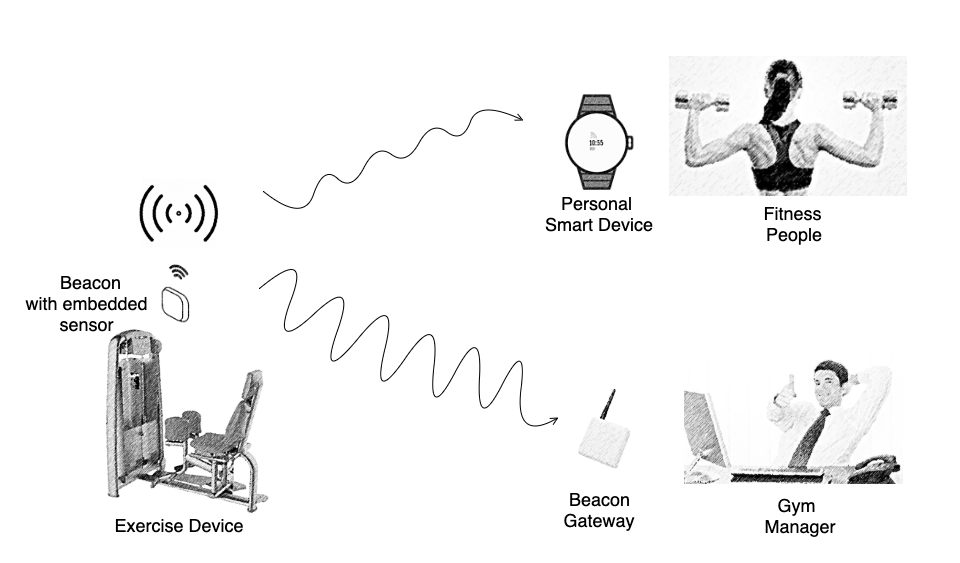}
\caption{System overview}
\label{structure}
\end{minipage}
\end{figure}


To reach the target of interest, three module components are needed for prototyping the ecosystem, as Figure \ref{structure} depicts. At first, a beacon embedded with a low-power inertial sensor working in advertisement mode is deployed on the moving part of the equipment, aiming to sense the working status of it. Second, a smartwatch or smartphone that can perceive the advertisement package on the user side is needed. Without manually recording the gym session, the user's smart device automatically records the gym activities by checking the received advertisement package that has the RSSI and equipment type and repetition information included. Third, a gateway is used to listen to all the beacon packages and send them to the cloud database, thus, a dashboard can be presented to the gym manager with real-time gym studio usage situation monitoring. The following subsections give a detailed description of each component used in the demonstrated ecosystem.

\subsection{Beacon and Gateway}

\begin{figure}[!t]
\begin{minipage}[t]{0.45\linewidth}
\centering
\includegraphics[width=\textwidth,height=3.0cm]{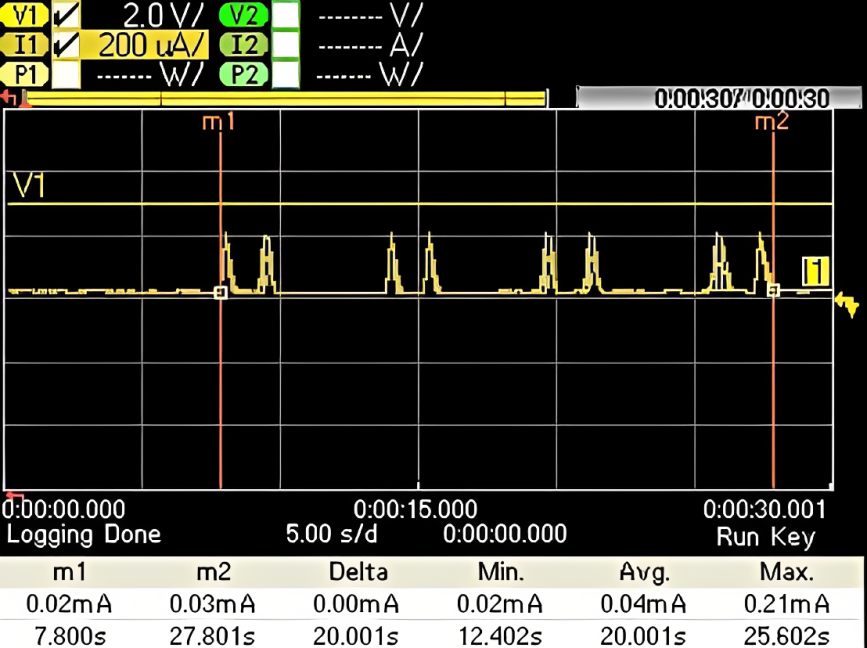}
\end{minipage}
\begin{minipage}[t]{0.45\linewidth}
\centering
\includegraphics[width=\textwidth,height=3.0cm]{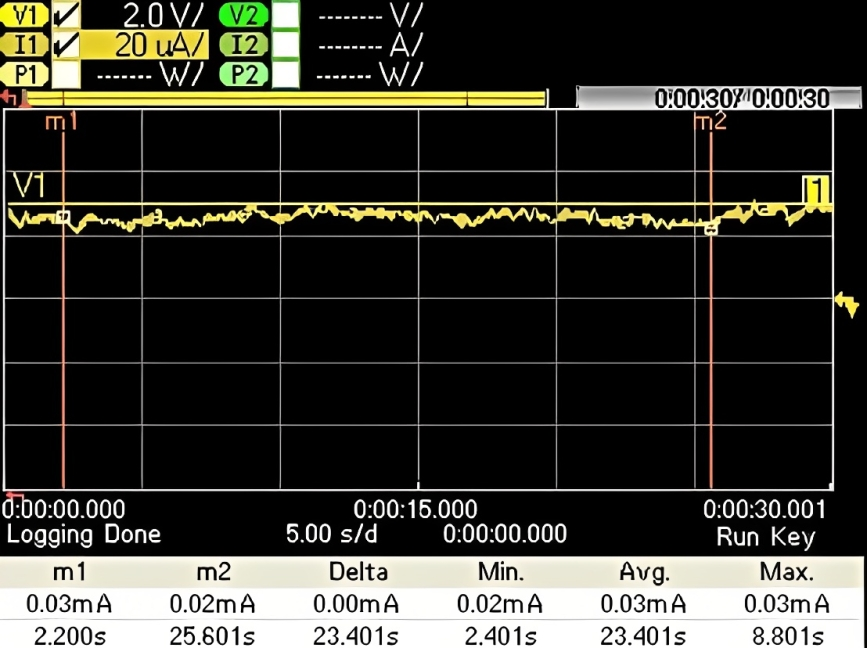}
\end{minipage}
\caption{System power consumption: advertising mode (left) and idle mode (right)}
\label{power}
\end{figure}

The heart of the beacon is an NRF52840 from Nordic Semiconductor, an advanced multi-protocol system-on-chip ideally suited for ultra-low power wireless applications. The embedded 2.4 GHz transceiver supports BLE-related functions like advertising and communication. The sensing component is the LIS3DH from STMicroelectronics, an ultra-low-power, high-performance three-axis linear accelerometer. The inertial sensor features ultra-low-power operational modes that allow advanced power saving and smart embedded functions. Specifically, the inertial sensor can be configured to generate interrupt signals using two independent programmable inertial events like wake-up and free-fall. By configuring the thresholds and timing of interrupt, the inertial sensor could work mostly in low power mode and generate an interrupt signal to NRF52840 only when a threshold is reached in a particular direction as defined. The similar working status is also located in NRF52840, as it works mainly in sleeping mode and changes to advertising mode only when an interrupt signal from the inertial unit is received. Figure \ref{power} presents how the power is consumed on the beacon measured by the Keysight N6705C DC power analyzer, showing an average current of the beacon with 0.03 mA in ideal mode and 0.04mA in advertising mode (four repetitions caused eight packages advertised in twenty seconds, we set double advertisements to decrease the loss rate of package perceiving), which indicates a battery life of around 400 days when powered by the CR2032 coin battery (210mAh, 80\% of the charge is regarded as efficient powering source), assuming that the beacon will work in advertising mode for six hours per day and idle mode in the left time. The iBeacon-form advertised package's universally unique identifier (UUID) chunk is pre-programmed to represent the gym equipment type. The minor data is updated for broadcasting the repetition number, which corresponds to the number of interrupts received from the inertial sensor. The repetition count is accumulated until no new interrupt signal is received in five seconds and reset to zero.

The gateway is an ESP32C3 SoC based on the open-source RISC-V architecture. We used it for its simplicity of programming and extensive WiFi and Bluetooth support. The availability of WiFi and Bluetooth 5.0 connectivity facilitates a variety of use cases based on dual connectivity. Especially the Bluetooth long-range support enables networking with great coverage and improved usability. To filter out any other Bluetooth advertisements, we stored each beacon identifier in the gateway beforehand so that it only focuses on the specific devices and ignores all other packages that are not advertised from the beacons. Using a permanently loaded repetition vector, the gateway keeps track of the current status of each machine. Each configured beacon and its correlating identifier is bound to a specific index in the array that gets updated with all received packages every couple of seconds. Every update also triggers the HTTP refreshing protocol. The gateway sends an updated HTTP PATCH packet to the Google database using an HTTP protocol. These packets consist of a JSON-form body with the repetition vector, the identifier of the respective machine, and an HTTP header that describes the sent information. The advertisement listening time is set to 3 seconds, and each HTTP packet transmission is measured with 0.9 seconds. For the gateway, the power consumption is not considered as it has no limitation of installing position and can be powered from a wire source.

\begin{figure}[!t]
\centering
\begin{minipage}[t]{0.80\linewidth}
\centering
\includegraphics[width=1.0\textwidth,height=4.2cm]{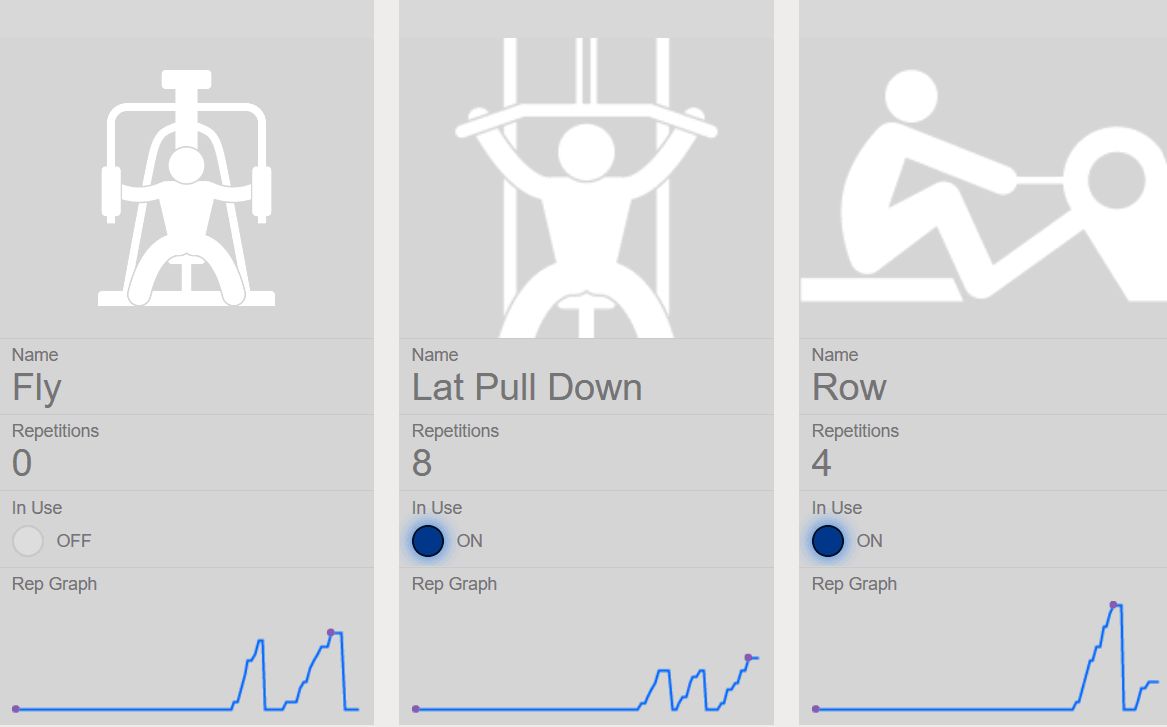}
\end{minipage}
\caption{Dashboard}
\label{dashboard}
\end{figure}

\begin{figure}[!t]
\centering
\begin{minipage}[t]{0.22\linewidth}
\centering
\includegraphics[width=\textwidth,height=3.7cm]{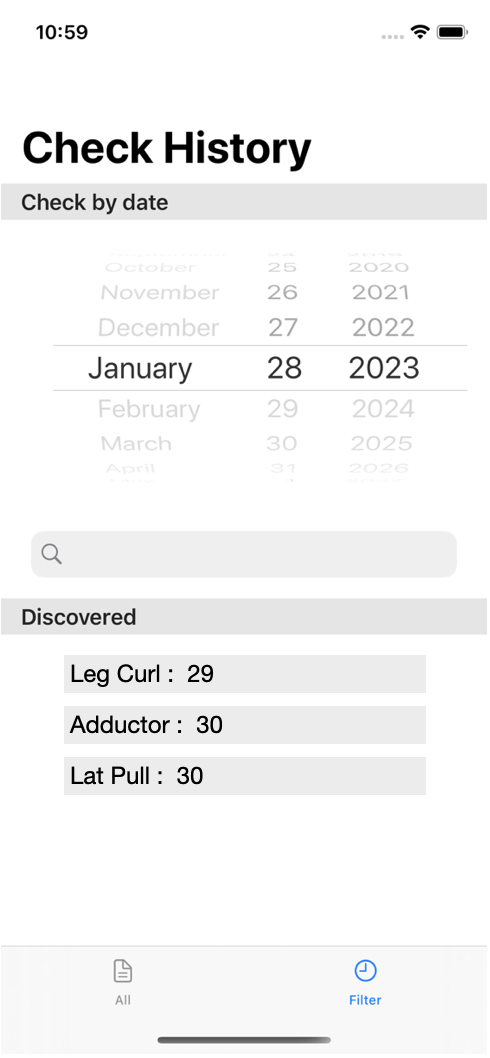}
\end{minipage}
\quad
\begin{minipage}[t]{0.22\linewidth}
\centering
\includegraphics[width=\textwidth,height=3.7cm]{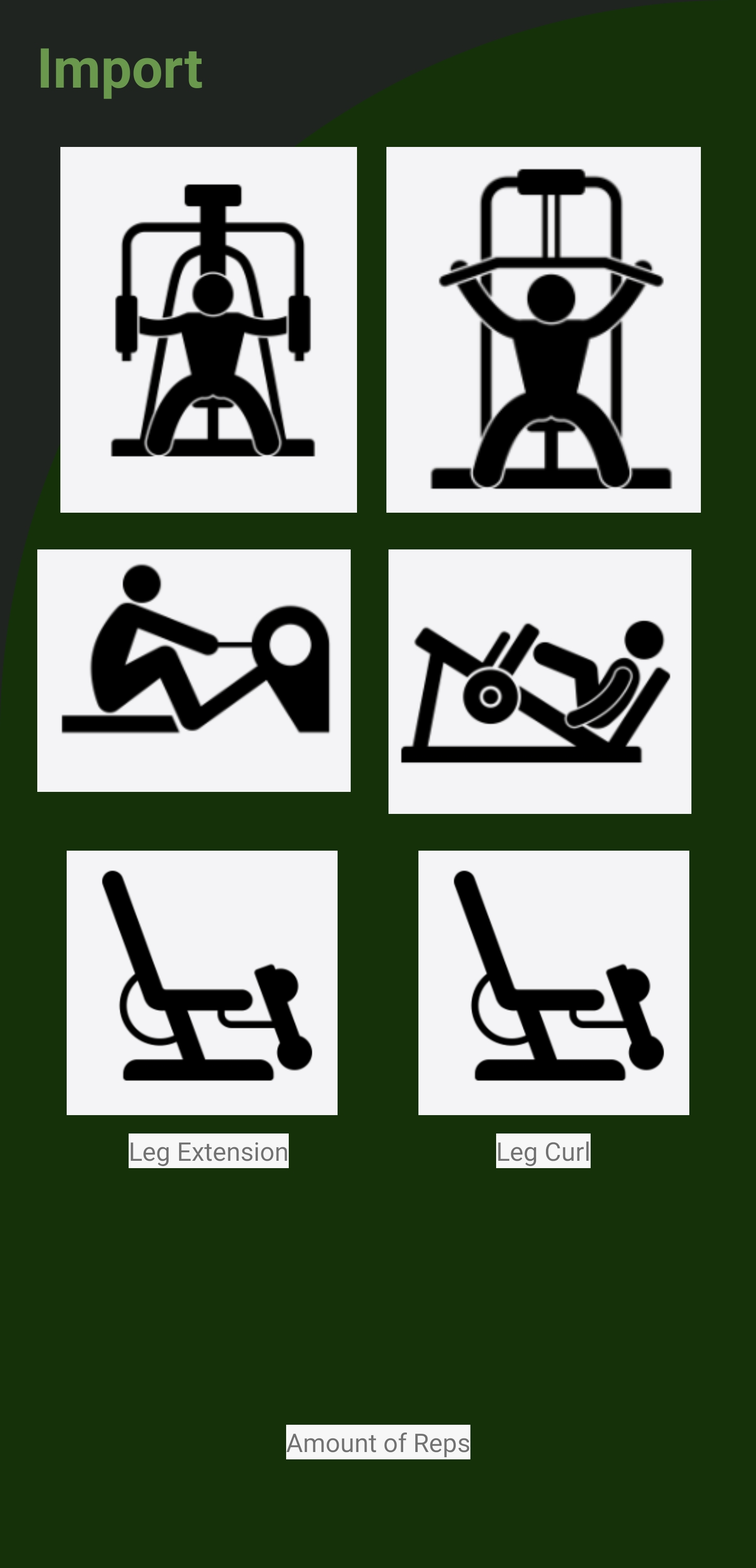}
\end{minipage}
\caption{Apps for user (left) and manager (right)}
\label{apps}
\end{figure}

\subsection{Smartwatch and Dashboard}

In the demonstration system, a programmable smartwatch is used by the gym user as the wearable device for gym session recording, and a dashboard is used for gym equipment status presentation. 

The smartwatch (LilyGo T-Wristband-NRF52 \cite{watch}) is commercially available and programmable. The watch features an NRF52832 SoC, an inertial measurement unit, a PCF8563 for Real-time clock/calendar, a capacitive touch button, a 160x80 resolution, general 0.96 inch LCD display Module (as Figure \ref{structure} right depicts), and a rechargeable lithium battery. A tiny programming board is delivered together for flashing. We programmed the firmware for the function switch by touching the capacitive button. As long as a "long-touch" occurs, the watch steps into Bluetooth scanning mode and perceives the nearby broadcasted packages. The RSSI value of each package is used to pick up the wearer's exercise, as a stronger RSSI means a closer distance from the beacon to the watch.

For the dashboard, we used an online service called Freeboard, developed by Bugs Labs, Inc., for real-time and interactive IoT visualization. The platform uses a simple GUI to stitch together panels in a grid layout. We used it to create vertical panels corresponding to each piece of equipment and displayed all available information (real-time and historical) related, as Figure \ref{dashboard} displays. 

Besides the smartwatch and dashboard, we also developed two smartphone apps for the user and the manager, as Figure \ref{apps} shows. The user one works similarly with the smartwatch. By activating the App, the smartphone's Bluetooth starts scanning the beacon packages and making the record. The manager one is connected to the firebase database, allowing internal logic and communication with the server. The two Apps aim to provide an alternative approach to data perception and presentation.

\section{Real-life experiment}

\begin{figure}[!t]
\centering
\begin{minipage}[t]{0.90\linewidth}
\centering
\includegraphics[width=0.95\textwidth,height=4.2cm]{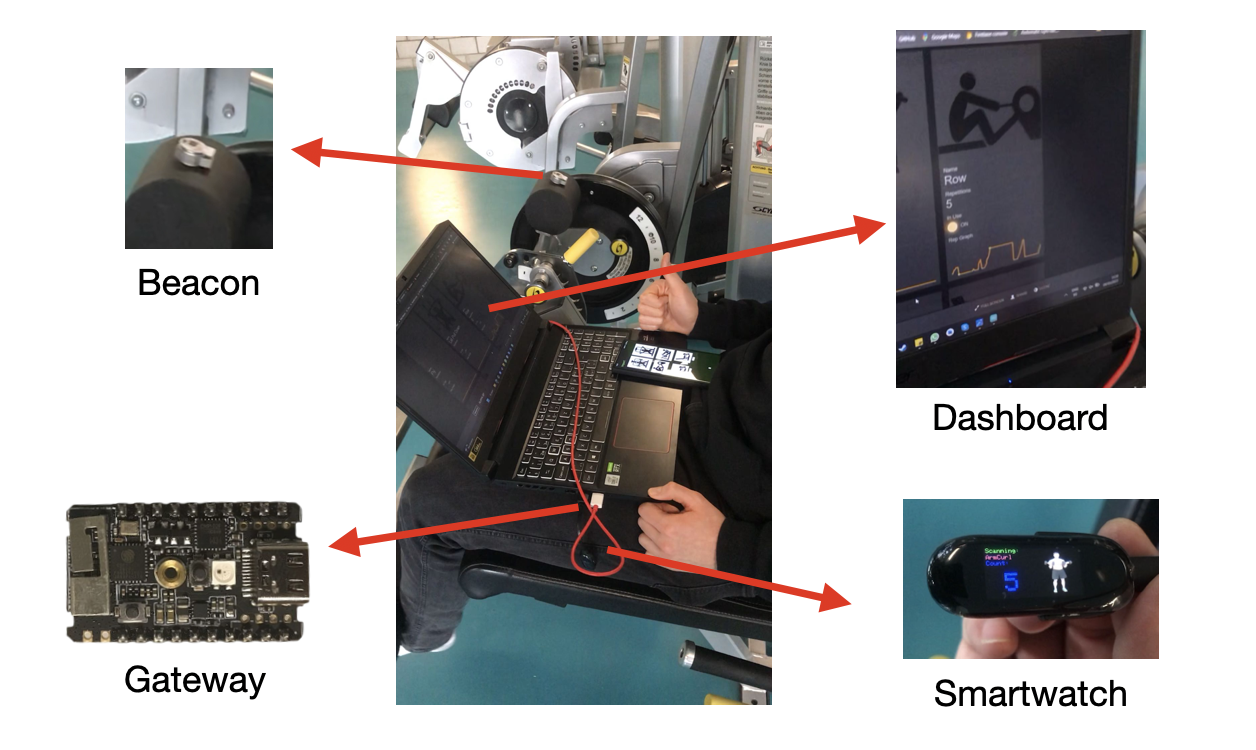}
\end{minipage}
\caption{Experiment with the crucial components in an university gym studio}
\label{experiment}
\end{figure}

To preliminarily test the system's feasibility, we conducted the experiment in a university gym studio. Figure \ref{experiment} is a screenshot of the recorded video. The gateway was connected to the in-house WiFi network and attached to our laptop to see the live debug feed. The beacon was attached to the part of the machines that undergoes the most movement during use(here, we first tested the equipment of leg-curl, leg-extension, and lat-pull). The readout from the dashboard and the smartwatch are used to verify the feasibility of the gym recording system. For the battery reason, the smartwatch is connected to the laptop but still close to the wrist. The exercise logos shown on the dashboard and smartwatch are not synchronized as they are not crucial to the test result during a preliminary experiment. We did three sets on each piece of equipment. Table \ref{result} lists the counting result. As can be seen, both the dashboard and smartwatch can present reliable repetition values most of the time. However, we noticed two imperfections in the preliminary test. First, the dashboard missed the last package quite often, such as set three on Leg-Curl and Leg-Extension, and set one on Lat-Pull. The reason behind locates in the advertisement scanning time of the gateway, as the gateway has to swing between WIFI uploading mode and Bluetooth scanning mode, which holds for 0.9 seconds and 3 seconds separately. Thus, the gateway was losing the advertisement package occasionally. As the repetition value is broadcasted directly from the beacon, the counting error caused by package loss will not be accumulated. Thus, we see it as a negligible flaw (or using more than one gateway to guarantee a full detection rate of the advertisement package). Second, during set three of Lat-Pull, the first six repetitions in the ten were successfully captured by the smartwatch and the gateway and showed in the dashboard. Still, as the beacon was not tightly attached to the moving frame of the equipment, a slight displacement caused the beacon to lose the orientation used for interrupt generation. This revealed a critical fact that the deployment site and direction of the beacon play an important role in reliable equipment status detection.

\begin{table}[!t]
\centering
\footnotesize
\caption{Preliminary gym recording test result (detected/truth)}
\label{result}
\begin{tabular}{|p{1.3cm}|p{0.6cm}|p{1.0cm}|p{1.0cm}|p{0.9cm}|p{0.7cm}|}
\hline
End Device & set & Leg-Curl & Leg-Extension & Lat Pull & Overall\\
\hline
\multirow{3}{*}{Dashboard} & set 1 & 10/10 & 10/10 & 9/10 & \\
\cline{2-5}
 & set 2 & 10/10 & 10/10 & 5/10 & 90.5\% \\
\cline{2-5}
 & set 3 & 9/10 & 9/10 & 5/5 &\\
 \hline
\multirow{3}{*}{Smartwatch}  & set 1 & Not activated & 10/10 & 10/10 &\\
\cline{2-5}
 & set 2 & 10/10 & 10/10 & 6/10 & 94.6\%\\
\cline{2-5}
 & set 3 & 10/10 & 10/10 & 5/5 & \\
\hline
\end{tabular}
\end{table}

\section{Future work}

The preliminary experiment demonstrates the feasibility of the proposed system for automatic gym exercise recording. However, a systematic and large-scale experiment still needs to be included to check the practical performance (we have only a few beacons and one smartwatch currently at hand). Thus, for future work, we will first make more critical hardware components for the following long-term and extensive experiment. Secondly, the smartwatch and dashboard are currently only for real-time data presentation. However, data storage space is needed for historical data playing. This function module on both the user and manager sides will be implemented in the following system versions. Thirdly, to cover more exercises other than the equipment-assistant ones, like the free weight exercises, this straightforward interrupt-triggered advertisement will be invalid because of the moving pattern complexity of the weights. Therefore, a precise data processing model will be explored to abstract the repetition information. Lastly, users may be interested not only in the repetition of an exercise but also in the weight loaded, which is out of the current system's ability. To address this, a separate sensing module is needed to provide extra exercise information.

\section{Conclusion}

This paper proposed an economical IoT solution for gym studio digitalization and automatic gym exercise recording. With low-power motion sensor embedded beacons, the usage status of each specific piece of gym equipment will be advertised through the beacon that features a battery life of around 400 days. Users' smart devices working in the advertisement listening mode will record the packages without interaction from the users. Thus the target of automatic gym exercise recording can be achieved. Meanwhile, a gateway collects all advertisements to monitor the gym equipment usage status. A dashboard enables the gym manager to have quick and visual access to all equipment and to make potential equipment upgrades or necessary machine additions. Depending on the daily usage, gym managers can make informed choices about their gym facility’s future. A preliminary real-life test with the critical system components shows the feasibility of the proposed gym exercise assistive system with 94.6\% accuracy, which can easily reach up to 100\% with a more suitable assembling. A long-term comprehensive experiment will be carried out in the future to demonstrate the practicability of the proposed solution.



\bibliography{IEEEabrv,sample}
\bibliographystyle{unsrt}

\end{document}